\documentstyle[12pt,epsf]{article} 
\newcommand{\be}{\begin{equation}} 
\newcommand{\en}{\end{equation}}
\newcommand{\bea}{\begin{eqnarray}}
\newcommand{\ena}{\end{eqnarray}}

\newcommand{\hbo}{\hbox to 1 true cm {\hfill } } 
\newcommand{\tr}{\hbox{tr}}

\newcommand{\fm}{\hbox{fm}}
\newcommand{\MeV}{\hbox{MeV}}
 
\begin{document} 
\vglue 1truecm
  
\vbox{UNITU-THEP-4/98 
\hfill \today 
}
  
\vfil
\centerline{\large\bf Interaction of confining vortices }
\centerline{\large\bf in SU(2) lattice gauge theory$^*$ } 
  
\bigskip
\centerline{M.~Engelhardt, K.~Langfeld, H.~Reinhardt and O.~Tennert}
\vspace{1 true cm} 
\centerline{ Institut f\"ur Theoretische Physik, Universit\"at 
   T\"ubingen }
\centerline{D--72076 T\"ubingen, Germany}
  
\vfil
\begin{abstract}

Center projection of SU(2) lattice gauge theory allows to isolate
magnetic vortices as confining configurations.
The vortex density scales according to the renormalization group,
implying that the vortices are physical objects rather than lattice 
artifacts. Here, the binary correlations between points at which
vortices pierce a given plane are investigated. 
We find an attractive interaction 
between the vortices. The correlations show the correct 
scaling behavior and are therefore physical. 
The range of the interaction is found to be $(0.4\pm 0.2) \, $fm, which
should be compared with the average planar vortex density of 
approximately $2 \, $vortices$/\fm ^2$. 
We comment on the implications of these results for recent discussions
of the Casimir scaling behavior of higher dimensional representation
Wilson loops in the vortex confinement picture.

\end{abstract}

\vfil
\hrule width 5truecm
\vskip .2truecm
\begin{quote} 
$^*$ Supported in part by DFG under contract Re 856/1--3. 
\end{quote}
\eject

\centerline{ \bf 1. Introduction \hfill }
\medskip 

Recently, the $Z(N)$ vortex picture of confinement has attracted
renewed interest \cite{tom93}-\cite{fa97}. Proposed as early
as 1978 \cite{aha78}, this picture assumes vortex type structures
to be responsible for the area law behavior of the Wilson loop.
These vortices, in the older literature also termed ``fluxons''
\cite{aha78},\cite{vinci}, each
contribute a factor $(-1)$ to the Wilson loop when they
pierce its minimal area. Fluctuations of the number of vortices
linked to a given Wilson loop produce a strong cancellation in its
expectation value, yielding the desired area law.

\vskip 0.3cm 
Subsequently, many theoretical as well as numerical efforts were devoted 
to identifying such vortex type confiners and elucidating their nature.
On the one hand, a glimpse of such configurations was afforded by the
``spaghetti vacuum'' \cite{ole} induced by the instability of
homogeneous chromomagnetic fields. On the other hand, efforts were
undertaken to define and classify vortex configurations comprehensively
on the lattice \cite{mack},\cite{brow}. A manifestly gauge invariant
description of vortices can be achieved by explicitly separating off
the center of the gauge group in the Yang-Mills link variables on
the lattice. In such a description, a distinction between ``thin''
and ``thick'' vortices \cite{tom97} arises. The abovementioned factor
$(-1)$ contribution to a pierced Wilson loop becomes the defining
gauge invariant property of a thick vortex.

\vskip 0.3cm
A fruitful approach to the investigation of specific infrared degrees
of freedom conjectured to be relevant for confinement was pioneered by
't Hooft \cite{tho76}. One utilizes the gauge freedom to bring an
arbitrary gauge field configuration as close as possible to the
type of configuration (``confiner'') under scrutiny; subsequently, 
one neglects residual deviations from the confiner (i.e. one projects
onto the latter) in the hope that the gauge fixing procedure has
concentrated onto it most of the relevant information contained in the
original gauge field configuration. The validity of this projection
procedure is difficult to establish a priori and it is more
commonly justified a posteriori by the success in reproducing, say,
the correct string tension.

\vskip 0.3cm
In this vein, 't Hooft introduced the so-called Abelian gauges, which
induce Abelian monopole singularities in the gauge-fixed fields.
Subsequently, only the Abelian monopoles are kept as relevant degrees
of freedom (so-called Abelian projection), allowing one to investigate
the possibility of confinement as a consequence of a dual Meissner
effect resulting from the condensation of the Abelian monopoles.

\vskip 0.3cm
In complete analogy, one can introduce so-called center gauges which
bring the link variables of a given lattice configuration as close as
possible to center elements of the gauge group \cite{deb96}. Vortices
are then defined and singled out by center projection (see below for
details). The crucial observation of the lattice calculations
\cite{deb96} is that a Wilson loop which is calculated with
center projected links gives rise to almost the full string tension
(a related conclusion is reached in the gauge invariant approach
\cite{tom97} mentioned further above). This implies that the center
gauge successfully concentrates the information relevant for
confinement onto the vortex degrees of freedom being projected on,
a state of affairs sometimes referred to as ``center dominance''.
By contrast, in quantities other than the Wilson loop, the generic error
due to the projection can be quite large~\cite{la97}. 

\vskip 0.3cm 
Any physical quantity $\rho $ (here, of mass dimension two) 
which is measured on the lattice in units 
of the lattice spacing $a$ must display a characteristic dependence on
the inverse gauge coupling constant $\beta = 1/g^2$, i.e. for 
sufficiently large $\beta $ and for a pure SU(2) gauge theory
\be 
\rho a^2 \; \approx \; const. \; \exp \left\{ 
- \frac{ 6 \pi ^2 }{11 } \, \beta \right\} \; \hbox to 5cm {\hfill 
(renormalization group) } \; . 
\label{eq:1} 
\en 
Any violation of the scaling law (\ref{eq:1}) signals that 
the field combination under examination is not a physical quantity.
Recently, some of us found \cite{la97} that the planar density $\rho $ 
of vortices piercing a given surface displays the 
desired scaling law (\ref{eq:1}), implying that the vortices originating 
from center projection are physical objects. In this letter, we
investigate this type of vortices. 

\vskip 0.3cm 
While the vortices defined by center projection are localized to
within one lattice spacing, it seems reasonable to assume that the
original unprojected gauge configurations associated with these
vortices are extended objects.
An important consequence of such a finite reach of the underlying
configurations is that the vortex vacuum may be able to correctly
describe the Casimir scaling behavior of higher dimensional
representation Wilson loops, contrary to earlier criticisms of
the vortex vacuum picture \cite{deb96},\cite{fa97}. For such a
mechanism to operate, one needs vortex diameters of one fermi
or more. In order to obtain some more information not least
concerning this point, we focus in this letter on the interaction
between the center vortices. A parallel incentive for such
an investigation lies in the observation (see below) that
the string tension is too small if 
correlations between vortices are neglected. We thus measure the binary 
correlation of vortex points piercing a given plane. 
We show that the correlation function is a physical quantity, 
since it scales according to the renormalization group equation (\ref{eq:1}). 
Our main result is that the vortex interaction is attractive and has 
a range of $(0.4\pm 0.2) \, $fm. The implications of our results 
concerning the Casimir scaling of Wilson loops in higher dimensional 
representations will be briefly addressed. 

\bigskip 
\centerline{ \bf 2. The random vortex vacuum \hfill }
\medskip 

Let us briefly review the definition of center vortices
introduced in~\cite{deb96}. For this purpose, a SU(2) link variable 
$U$ is decomposed as 
\be 
U \; = \; \alpha _0 \; + \; i \vec{\alpha }
\; \vec{\sigma } \; , \hbo 
\alpha _0^2 + \vec{\alpha }^2 =1 \; . 
\label{eq:1b}
\en 
In the Abelian gauge, the magnitudes of the so-called charged components 
$\alpha _1$, $\alpha _2$ are minimized with the help of
gauge transformations; specifically, one maximizes 
$\sum_{i} \tr \, (U_i \sigma^{3} U_i^{\dagger } \sigma^{3} )$,
where $i$ is a superindex labeling all the different links on the
lattice. The Abelian projected links $U^A $ 
are then defined by disregarding the charged components, i.e. 
\be 
\hbox to 4 cm{ \hfill Abelian projection: \hfill } U 
\rightarrow U^A \; = \; \frac{ \alpha ^\prime _0 \; + \; i 
\alpha ^\prime _3 \; \sigma^{3} }{ \sqrt{ \alpha _0^{\prime \, 2 } 
+ \alpha _3^{\prime \, 2} } }  \; . 
\label{eq:2}
\en 
The Abelian gauge still allows for U(1) gauge transformations of the type 
$\exp (i \eta \sigma^{3} ) $. The center gauge fixes the residual gauge 
degree of freedom by demanding that the residual gauge transformation
maximize $\sum_{i} (\tr \, U_i^A )^2 $. After adopting the center gauge, 
center projection is defined by disregarding the 3-component, i.e. 
\be 
\hbox to 4 cm{ \hfill Center projection: \hfill } U^A
\rightarrow U^C \; = \; \alpha ^{\prime \prime }_0 / \vert 
\alpha ^{\prime \prime }_0 \vert 
\; \in \; \{ \pm 1 \} \; . 
\label{eq:3} 
\en 
A plaquette on the lattice is defined to be part of a (center) vortex, 
if the product of the center projected links which span the plaquette 
under consideration yields $-1$. A visualization of these points 
(for a given time slice) shows that these points are indeed grouped 
to string-like objects~\cite{la97}. 

\vskip 0.3cm
The crucial result of \cite{deb96} was the observation that
one recovers almost the full string tension when calculating
the Wilson loop with center projected links instead of 
using full link variables. It should be mentioned that center
projection can also be performed after a direct maximal center
gauge fixing without preceding Abelian projection. In this case,
the string tension agrees even better with the full one \cite{deb96}.
Center projection evidently does not 
strongly truncate the infrared degrees of freedom which are 
responsible for confinement (center dominance).
Resorting to the Stokes theorem, one easily sees that the product of 
center projected links which lie on the circumference of a Wilson area 
yields $(-1)^n$ if $n$ denotes the number of (center) vortices which pierce 
this area~\cite{la97}. Hence, a vacuum consisting of (center) vortices 
reproduces, via the relation
\be
\langle W[{\cal C}] \rangle
\; = \; \sum _{n=0}^{\infty } (-1)^n \, P(n) \; ,
\label{eq:4}
\en
the approximate expectation value of the Wilson 
loop obtained with center-projected links,
where $C$ is the Wilson loop, and $P(n)$ is the probability that $n$ 
vortices pierce its minimal surface. 

\vskip 0.3cm
Let us now neglect correlations between the vortices and calculate the
string tension obtained in such a random vortex vacuum.
Assume that the lattice volume is $L^4$, whereas the minimal surface of 
the Wilson loop under consideration possesses the area 
${\cal A}$, ${\cal A} \ll L^2$. The Wilson loop is embedded in a 
plane $H$ of area $L^2$. The random vortex model assumes that the 
probability $p$ that a vortex which pierces $H$ also pierces the
Wilson area is $p= {\cal A}/L^2$. If $N$ vortices pierce the plane 
$H$, then the probability $P_{rand} (n)$ that precisely $n \; (\le N)$ 
vortices pierce the Wilson area is 
\be 
P_{rand} (n) \; = \; \left( \begin{array}{c} N \\ n 
\end{array} \right) \; p^n \; (1-p)^{N-n} \; . 
\label{eq:5} 
\en 
Hence, the expectation value (\ref{eq:4}) in the random vortex model gives
\be 
\langle W[{\cal C}] \rangle _{rand} \; = \; \sum_{n=0}^{\infty }
(-1)^n P_{rand} (n) \; = \;
(1 \; - \; 2 \, p) ^{N}\; . 
\label{eq:6} 
\en
Our lattice simulations~\cite{la97} revealed that the planar vortex
density $\rho = N/ L^2 \approx 2 / \fm ^2 $ is a physical quantity,
where the string tension $\kappa = (440 \, \MeV )^2 $ was used to fix
the renormalization scale.
We therefore obtain in the infinite volume limit ($L \rightarrow \infty$, 
i.e. $N \rightarrow \infty $, $\rho $ fixed) 
\be 
\langle W[{\cal C}] \rangle _{rand} \; = \; \lim _{N \to 
\infty } \left( 1 \, - \, \frac{2 \rho {\cal A} }{N} 
\right)^{N} 
\; = \; \exp \left( - 2 \, \rho \, {\cal A} \right) \; . 
\label{eq:7} 
\en 
Eq.(\ref{eq:7}) yields the desired area law, from which we read off 
the string tension 
\be 
\kappa _{rand} \; = \; 2 \, \rho \; \approx \; (400 \, \MeV )^2 \; , 
\label{eq:8} 
\en 
which should be compared with the exact (input) value 
$\kappa = (440 \, \MeV )^2 $. While inserting
the exact probability distribution $P(n)$ generated in our lattice
simulations into (\ref{eq:4}) yields 
almost the full string tension, the value in the random vortex
vacuum turns out to be 17\% too small. The correlations between the 
vortices are obviously significant. 

\bigskip 
\centerline{ \bf 3. Vortex correlations \hfill }
\medskip 

In order to study the correlations between the vortex points on the 
plane $H$, we introduce a field $s_j $, where $j$ is a superindex
labeling all the different plaquettes in the lattice: $s_j $ is $1$, if 
plaquette $j$ is part of a vortex, and is $0$ otherwise.
The lattice average of $s_j $ 
is independent of $j$ due to homogeneity and
isotropy. It is directly related to the vortex density, i.e. 
\be 
\rho \, a^2 \; = \; \langle s_j \rangle  \; . 
\label{eq:9} 
\en 
Consider next the normalized correlator
\be
c_{ij} \; = \; \frac{\langle s_i s_j \rangle }{\langle s_i \rangle
\langle s_j \rangle } \; ,
\label{cijdef}
\en
where in the following, plaquette $i$ and plaquette $j$ will be
considered to lie in the same plane, $l$ lattice spacings
(i.e. a distance $r=l\cdot a$) apart in the direction of one of the
coordinate axes. Due to homogeneity and isotropy, the corresponding
$c_{ij} $ will only depend on $r$ and will thus henceforth be denoted
as $c(r)$. This correlator has a very transparent interpretation in
terms of a conditional probability: Assuming that one sits on a
plaquette which is part of a vortex, $\rho a^2 c(r) $ is the 
probability of finding another vortex at a distance $r$. This is
precisely the algorithm used in practice to extract $c(r)$. The
quantity $c(r)$ is normalized such as to give a constant equal to unity
if the vortices piercing the plane under consideration are
statistically independent; thus, $c(r)$ constitutes what is often
termed a (planar) radial distribution function. Interactions produce
deviations from unity; the distance scales over which the deviations
persist give a rough estimate of the range of the interaction in the
medium. 

\vskip 0.3cm 
Since the vortices are physical objects, their interaction as 
revealed in $c(r)$ should also behave as a physical quantity under
the renormalization group. In order to verify this, it is necessary
to examine the dimensionless function $c(r)$ at different couplings
$\beta $, where it is crucial to take into account the running of
the lattice spacing $a(\beta )$ entering the physical distance
$r=l\cdot a$. In order to estimate the statistical errors as well as the 
influence of systematic errors, we used three methods to extract 
$a(\beta )$ in physical units. Firstly, 
we measured the string tension $\kappa a^2(\beta )$ for $\beta $ 
values within the scaling window $2 < \beta < 2.8 $. Using $\kappa =
(440 \, \MeV )^2 $, this procedure directly yields $a(\beta )$ in physical 
units. Secondly, we fitted the perturbative scaling law (\ref{eq:1}) to 
$\kappa a^2 (\beta )$, and used the formula 
(\ref{eq:1}) to express $l\cdot a=r$ 
in physical units (``ideal scaling''). Thirdly, we extracted the ``running'' 
of $a(\beta )$ from the measured quantity $\rho a^2 (\beta )$, and used
$\rho \approx 2 \, \fm ^{-2}$ as physical reference scale. 

\begin{center} 
\begin{tabular}{|ll|} \hline 
measured $\kappa \, a^2 \; \; \rightarrow \; \; a(\beta ) $ & 
``measured scaling'' \\ 
fit of $\kappa \, a^2 $ to (\ref{eq:1}) & 
``ideal scaling'' \\ 
measured $\rho \, a^2 \; \; \rightarrow \; \; a(\beta ) $ & 
``density scaling'' \\ \hline 
\end{tabular} 
\end{center}

\begin{figure}[t]
\centerline{ 
\epsfxsize=7cm
\epsffile{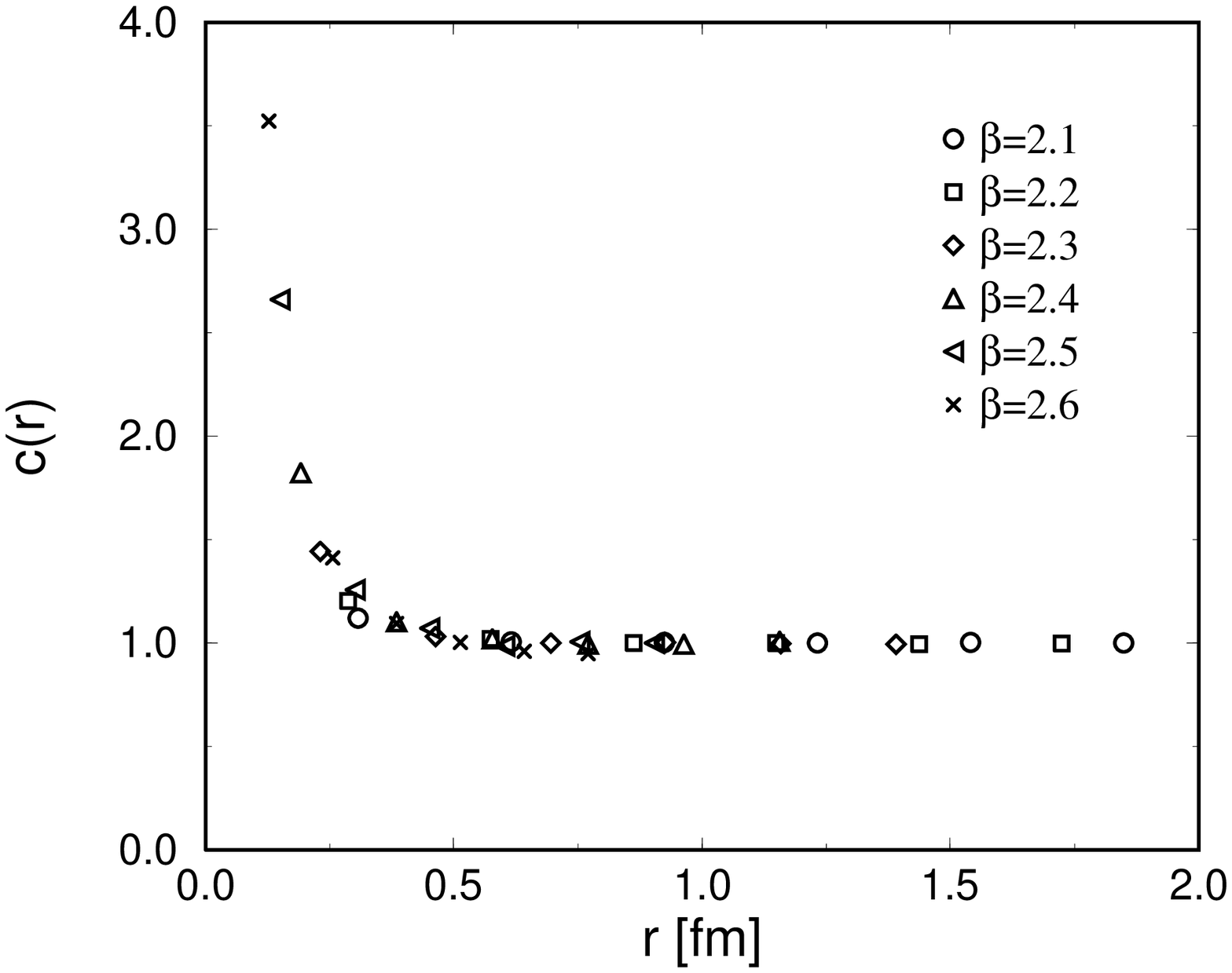}
\epsfxsize=7cm
\epsffile{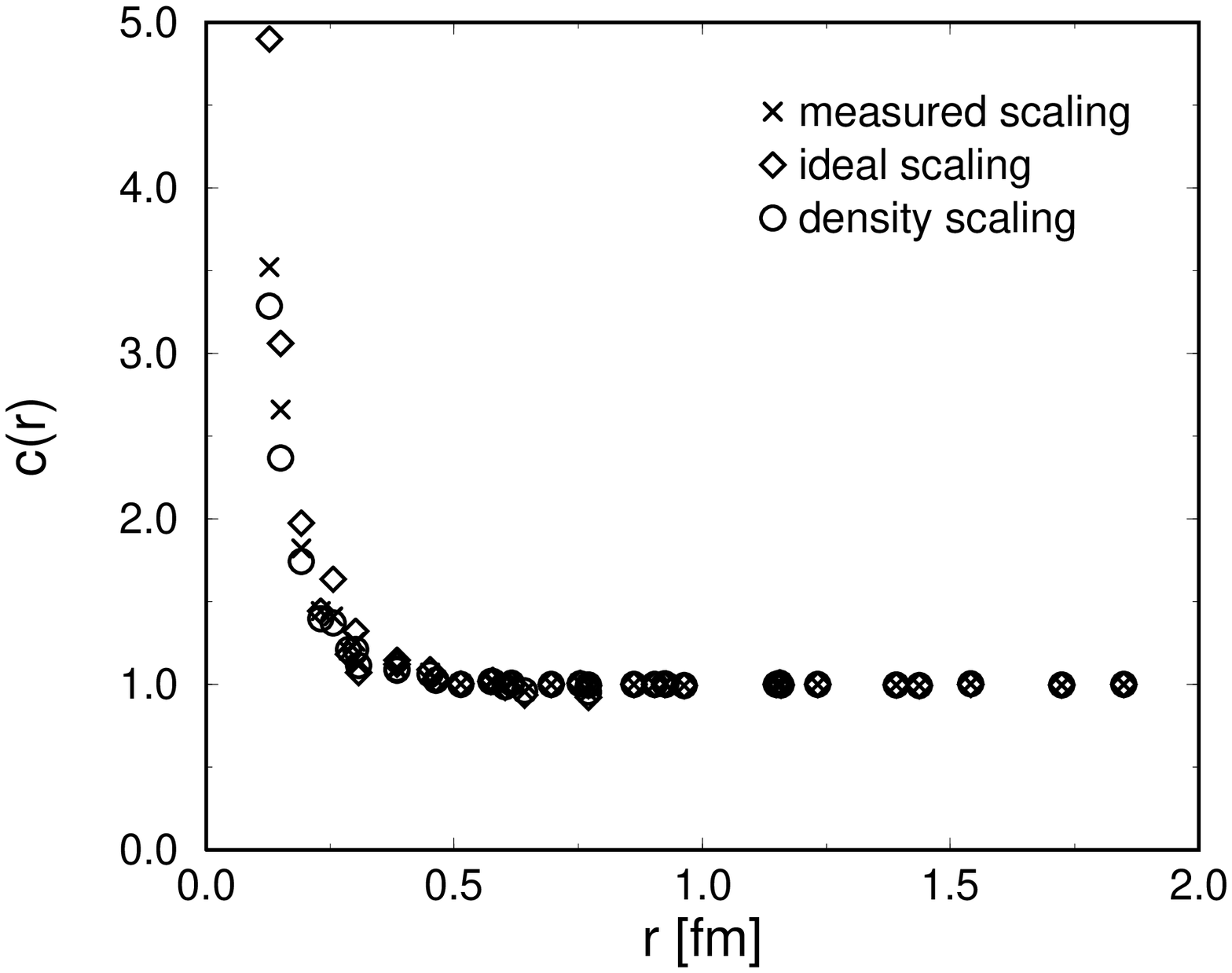}
}
\vspace{-.8cm} 
\caption{The planar radial distribution function $c(r)$ for points
at which vortices pierce a given plane.}
\label{fig:1} 
\end{figure} 

Within the statistical error bars, all three methods of extrapolating
to the continuum limit should yield the same results. 
Figure \ref{fig:1} shows our numerical results for the planar radial
distribution function $c(r)$ as a function of $r$.
We have used lattices consisting of 
$10^4$ and $12^4$ lattice points in order to estimate the finite 
size effects. Calculations with both lattice sizes yield the same results 
within the statistical errors. In the left hand picture, the extrapolation 
of the data was done with ``ideal scaling''. The crucial observation is that 
the result is indeed renormalization group invariant, i.e. independent of
the actual choice of $\beta $. Consequently, $c(r)$ is a physical quantity. 
We further corroborate this with the right hand picture, in which the
different types of scaling mentioned above are confronted with each
other, for a $10^4 $ lattice.

\vskip 0.3cm 
The shape of the planar radial distribution function $c(r)$ plotted in
Figure \ref{fig:1} reveals that an attractive interaction operates
between the vortices in the vortex medium. Note that the range of this
interaction constitutes a rather vaguely defined notion. One way of
defining the range of an attractive interaction is to look for the
first crossover of the radial distribution function below unity (note
that such a crossover must exist, since an appropriate integral over
the radial distribution function must reproduce the total number of
vortex points). This crossover happens for the present data at
$r=0.6 \, $fm, where it must be noted that in this region the statistical
errors are already of the same magnitude as the deviation from unity.
The value $r=0.6 \, $fm can be regarded as an upper limit on the range
of the interaction. Another possible definition of the interaction
range lies in fitting an exponential decay to the deviation of the
radial distribution function from unity and thus extracting a typical
screening length. This yields a value of roughly $0.2 \, $fm, which can
be regarded as a lower limit on the interaction range.
There is an intuitive argument making the appearance of this
scale plausible: The planar radial distribution function is by its
definition (\ref{cijdef}) nothing but a plaquette-plaquette
correlation function, albeit with center-projected links, and shifted
by unity. The exponential fall-off of such correlation functions
is generically controlled by glueball masses. Thus, the
emergence of the relatively high energy scale associated with a
screening length of $0.2$ fm in $c(r)$ is not too surprising.

\vskip 0.3cm
Finally, it should be noted that the planar correlations measured here 
still represent a rather unspecific yardstick for the structure of the
vortex vacuum. They subsume a variety of more detailed effects; not only
are they sensitive to the actual interaction of segments of neighboring
vortices, but also e.g. to the shape distributions of the individual
vortices in the directions orthogonal to the plane under consideration.
It would be interesting to 
further disentangle the effects of the actual vortex-vortex interaction
and the effects due to, say, curvature terms in the single-vortex
action.

\bigskip 
\centerline{ \bf 4. Conclusions \hfill }
\medskip 

We have further investigated the center vortices introduced
in~\cite{deb96}. These vortices can account 
for almost the full string tension, and were recently recognized as 
physical objects (rather than lattice artifacts)~\cite{la97}, since 
the density $\rho $ of vortices piercing a plane 
scales according the renormalization group. 

\vskip 0.3cm 
Here, we have observed that the random vortex model, in which
correlations between the vortices are neglected, qualitatively
describes the gross features of confinement, but underestimates
the string tension by 17\% . In order to study the binary vortex
correlations induced by the full Yang-Mills action, we have
introduced the planar radial distribution function $c(r)$.
Our numerical simulations reveal that $c(r)$ is a 
renormalization group invariant and therefore a physical quantity. 
The data show that the vortex interaction is attractive and possesses 
a range of $ (0.4\pm 0.2) \, $fm in the vortex medium.  

\vskip 0.3cm 
With regard to the possibility of correctly describing the Casimir
scaling of Wilson loops in higher dimensional representations along the
lines discussed in \cite{fa97}, our results do not allow very
definite conclusions. If we roughly identify the range of the
vortex-vortex interaction, as read off from the measured planar
radial distribution function, with the diameter of the {\em unprojected}
gauge configurations associated with the center vortices, we
reach the conclusion that these configurations are on the average
too thin to allow for the mechanism of Casimir scaling discussed
in \cite{fa97}. However, it is entirely possible that the vortices
are significantly thicker and through some cancellation only start
to feel an appreciable attractive interaction when they already
considerably overlap. Therefore, our present results do not 
necessarily contradict the mechanism of Casimir scaling of Wilson 
loops in higher dimensional representations proposed in \cite{fa97}.

\end{document}